\documentclass[11pt,reqno]{article}
\usepackage{amsfonts}
\usepackage{amsmath}
\usepackage{amsbsy}
\usepackage{graphicx}
\usepackage{amssymb,latexsym}
\usepackage{graphicx}
\usepackage{caption}
\usepackage{subcaption}
\numberwithin{equation}{section}

\begin{document}

\begin{titlepage}

\title{Revisiting Lifshitz-type solutions in $R^2$-corrected gravity}

\author{Se\c{c}il \c{S}entorun\footnote{E.mail:secilo@eskisehir.edu.tr} \\
{\small Department of Physics, Eskisehir Technical University, 26470 Eski\c{s}ehir, Turkey}}

\date{ }

\maketitle

\bigskip

\begin{abstract}

\noindent In this study, we construct exact higher-dimensional Lifshitz-type solutions in $R^2$-corrected gravity at the critical point of the theory, where the field equations become degenerate because of the vanishing of the effective gravitational coupling. The analysis is performed on product manifolds of the form  $Li_{m}\times \Omega_{(n-m)}$, where $Li_{m}$ denotes an $m$-dimensional Lifshitz-type spacetime exhibiting anisotropic scaling with dynamical exponent $z$ and $\Omega_{(n-m)}$ represents an $(n-m)$-dimensional space of constant curvature. This geometric decomposition allows for a unified treatment of static, stationary (rotating), and hyperscaling-violating configurations within a purely gravitational framework.

We show that the theory admits new broad families of exact Lifshitz black hole and black brane solutions, including extremal configurations, whose scaling exponents and horizon structures are constrained by higher-curvature terms. The stationary solutions are interpreted as rotating Lifshitz-type black holes with well-defined Killing horizons within the appropriate parameter ranges. Owing to the critical nature of the theory, these solutions exhibit vanishing entropy and zero conserved charges despite having non-zero temperature, reflecting the degenerate structure of the field equations. Our results extend the previously known Lifshitz constructions in Einstein and higher-derivative gravity and provide a systematic higher-dimensional framework for exploring anisotropic and hyperscaling-violating geometries supported by curvature-squared interactions.

\end{abstract}

\end{titlepage}

\section{Introduction}

General Relativity (GR) is the most concise and effective theory of gravity developed in the past century, as it is thought to describe the macroscopic features of spacetime. Although GR has had a significant impact on the scientific community, it is necessary to make both large- and small-scale modifications to GR. Large-scale changes may be indicated by various astrophysical observations and potential departures from geodesic motion caused by gravitational self-force. However, a short-range adjustment is required for renormalizable, unitary quantum field theory of massless spin-2 gravitons. It is crucial to consider both ultraviolet (UV) and infrared (IV) complete theories of gravitation when examining scale-covariant generalizations of gravitational theories \cite{weyl, eddington}. Because of these theoretical limitations, several modifications have been suggested. Of these modifications, the most intriguing is the $f(R)$ gravity model, where the gravitational Lagrangian is described by a function of the curvature scalar $R$. $f(R)$ gravity theory allows for a broader class of gravitational theories that can include effects such as dark energy (a hypothetical form of energy proposed to explain the accelerated expansion of the universe), dark matter (a substance that shapes the cosmos) and inflation (the period of accelerated expansion in the early universe) without the need for additional fields  \cite{sotiriu, felice}. There has been interest in f(R) theories of gravity because they can have improved renormalization properties \cite{stelle}, can lead to an inflation period \cite{starobinsky}, and result from the quantization of matter fields in an unquantized space-time  \cite{utiyama}. Recently, they have attracted a lot of interest as a possible explanation for the observed late-time accelerated expansion of the universe \cite{nojiri} and as a testing tool to compare the predictions of $ f(R)$ gravity models against observations coming from observations of the universe's structure \cite{clifton}. Modified gravity theories have been extensively studied as a robust framework for explaining the evolution of the universe. In particular, comprehensive reviews \cite{nojiri1,nojiri2} provide a unified description of cosmic history, covering everything from early inflation and bouncing to the late acceleration phase. Considering all this, $f(R)$ gravity theory is significant because it provides a natural inflationary scenario, is testable against observations, and provides an alternative to dark energy and dark matter.

\noindent $f(R)$ theories can either be studied in pseudo-Riemannian spacetimes, where the connection is constrained to be Levi-Civita (torsion-free), or they can be modeled in Riemann-Cartan spacetimes involving spacetime torsion. To derive the field equations of $f(R)$ gravity theory, these two distinct approaches can be employed, where the former can also be identified as metric formalism, and the latter can be named as Palatini formalism \cite{palatini}. In metric formalism (thoroughly examined in \cite{buchdahl}), the field equations are derived by varying the metric tensor considering that the connection is Levi-Civita (without torsion), such that the connection depends on the spacetime metric. On the other hand, Palatini formalism can be employed by considering metric and connection fields as independent variables when varying the action. Although these two approaches yield identical field equations for the GR, they differ significantly in the modified gravity models. In particular, for the case in which the gravitational Lagrangian involves higher-order curvature terms, independent variations of the Lagrangian yield field equations that involve propagating torsion. 

\noindent A particular extension of GR can be provided by quadratic curvature gravity models that include $ R^{2} $, Riemann -curvature, and Ricci -curvature squared terms \cite{stelle, buchdahl, lanczos}. In cosmological scenarios, such models can be viewed as viable gravity theories that explain the accelerating expansion of the universe \cite{nojiri3}. The physical meaning of the theory was investigated by focusing on fundamental concepts such as gravitational energy, momentum, and mass of theory \cite{deser}. Exact solutions for plane-fronted gravitational waves in an Anti-de Sitter (AdS) background are provided in \cite{gullu}, Kerr-Schild type metrics are explored, and new solutions for AdS waves and spherical waves in an AdS background are given in \cite{gurses}. It is also of great interest to obtain black hole solutions and analyze the properties and existence of the black hole solutions in these gravity models \cite{lu,kehagias}. 

\noindent  Gauge/gravity correspondence offers a fascinating framework for examining strongly coupled quantum field theories by connecting them to gravitational theories in higher dimensions. Lifshitz spacetimes, first introduced in \cite{kachru}, serve as important structures for understanding non-relativistic scaling behaviors that deviate from the conformal symmetry observed in AdS spacetimes. It has been demonstrated that three-dimensional massive gravity admits Lifshitz metrics as exact solutions featuring a dynamical exponent $z$. Crucially, for $z = 3$, it yields exact analytic black hole solutions that asymptotically serve as lower-dimensional analogs of gravity duals for scale-invariant fixed points \cite{ayonbeato3}. The generalized work on a four-dimensional $R^2$-corrected Lifshitz black hole (with $z = 3/2$) \cite{cai} into a two-parameter family that works for any dynamical exponent $z$ and arbitrary dimension $D$ can be found in \cite{ayonbeato2}. A notable outcome is the identification of an extremal Lifshitz black hole under a specific parameter relation. The analysis also reveals solutions with logarithmic fall-off for critical $z$, and, in dimensions $D \geq 5$, they present three distinct families of analytic Lifshitz black holes valid for general $z$. Among these, one family includes, in limiting cases, both the $z = 3$ three-dimensional Lifshitz black hole and a new $z = 6$ four-dimensional black hole-underlining the continuity across dimensions and exponents. Charged Lifshitz black holes in the context of Einstein–Proca–Maxwell systems for any $z \geq 1$, with a critical exponent $z = D – 2$ emerged as a special case that asymptotically produced Lifshitz black holes with logarithmic decay, supported by logarithmic electrodynamics. The solution space also includes extremal black holes, and charged topological Lifshitz black holes via slight generalizations \cite{alvarez}. Although the existing literature contains limited exact analytical solutions for Lifshitz black holes, notable contributions, such as the work of \cite{sarioglu}, have identified stationary D-dimensional spacetimes that qualify as Lifshitz black holes under specific conditions. 

\noindent The study of Lifshitz spacetimes remains a cornerstone in exploring the gravitational duals of non-relativistic quantum field theories, particularly those arising in condensed matter systems, which were initiated by seminal works generalizing the AdS/CFT correspondence to non-relativistic contexts with anisotropic scaling \cite{hartnoll}. A critical area of research involves characterizing the thermodynamic and microscopic properties of Lifshitz black holes. In this regard, key efforts have been made to establish microscopic foundations of black hole entropy in these geometries. For example, the work of \cite{ayonbeato3} concentrated on the microscopic entropy of higher-dimensional nonminimally dressed Lifshitz black holes, providing insight into the underlying degrees of freedom. Continuing in this thermodynamic vein, \cite{bravogaete} considered the Cardy entropy of charged rotating asymptotically rotatingAdS and Lifshitz solutions, including a generalized Chern-Simons term. In this study, it is shown that higher-order derivative terms have a direct influence on the statistical mechanics and entropy relations of anisotropic solutions.

\noindent This study aims to expand the understanding of Lifshitz black holes by seeking higher-dimensional new exact analytic solutions that satisfy these relationships for both static and stationary cases. To this end, we investigate Lifshitz-type static and stationary solutions in $R^{2}$-corrected gravity theory, which involves the usual Einstein-Hilbert extended with $R^{2}$ term and a cosmological parameter. Motivated by the work of \cite{gregory}, which investigates Lifshitz-type product manifold solutions in supergravity, we construct $Li_{m} \times \Omega_{(n-m)}$ type solutions in $ R^{2} $-corrected gravity theory where $ Li_{m} $ corresponds to an $m$-dimensional Lifshitz-type subspace, and $ \Omega_{(n-m)} $ describes the $(n-m)$-dimensional subspace. We obtained both static and stationary solutions by investigating the thermodynamic properties of the solutions. In addition, we explicitly illustrate that this theory admits hyperscaling violating- type solutions. In contrast to existing constructions that focus on specific dimensions or special values of the dynamical exponent, the present study provides a unified higher-dimensional framework covering static, stationary, and hyperscaling-violating Lifshitz-type geometries within the same $R^2$-corrected gravity setup. Within this setup, we identify the conditions under which constant scalar curvature solutions exist and analyze their geometric structures. Although Lifshitz-type solutions in modified gravity have been extensively studied, the present work aims to extend these configurations within a quadratic curvature framework in a systematic manner. In particular, we focused on constructing higher-dimensional solutions and examining their thermodynamic properties in a unified setting. Such generalizations are of interest in the context of holography, where Lifshitz spacetimes provide gravitational duals to non-relativistic field theories with anisotropic scaling \cite{taylor}.

\noindent The structure of this paper is as follows. In Section 2, we provide mathematical preliminaries and clarify the notation. In Section 3, we present the action and variational procedures used to derive the field equations of the $R^2$-corrected gravity theory. In Section 4, we construct Lifshitz-type static and stationary solutions in higher dimensions. In the same section, we discuss some thermodynamic properties of the solutions, including extremal cases. Finally, we conclude the paper in Section 5.

\section{Mathematical Preliminaries}

The language of exterior differential forms was employed throughout this work. We adopt both Latin and Greek index conventions: Latin indices $(a,b,c,\dots)$ denote orthonormal frame components, whereas Greek indices $(\mu,\nu,\dots)$ refer to coordinate components on the spacetime manifold. The Hodge dual is represented by $\ast$ and the wedge product by $\wedge$. Unless otherwise stated, the wedge product between differential forms is assumed to be implicit. For consistency, repeated indices imply a summation following the Einstein convention.

\noindent The spacetime metric tensor is given by $g=g_{\alpha \beta} dx^{\alpha}\otimes dx^{\beta}=\eta_{ab} e^a \otimes e^b$ where $\{e^a\}$ and $\{dx^{\alpha} \}$ represent the orthonormal co-frame 1-forms and the coordinate basis 1-forms, respectively. Here, $\eta_{ab}=diag (- + ... +)$ are the orthonormal components and $g_{\alpha \beta}$ are the coordinate components of the metric. The orthonormal basis of frame vectors $\{ X_a\}$ is dual to the orthonormal co-frame 1-forms, satisfying $e^b (X_a)=\iota_a e^b={\delta_a}^b$, where $\iota_a$ denotes the interior product operation with respect to frame vectors $X_a$. The oriented volume form is $\ast 1=e^0 \wedge e^1 \wedge ... \wedge e^n$. In this study, the connection structure of spacetime is assumed to be metric-compatible, that is, $Dg=0$ or $D\eta_{ab}=\omega_{ab}+\omega_{ba}=0$. Here, $D$ denotes the covariant exterior derivative operator acting on the form fields, and $\{{\omega^a}_b\}$ represents the connection 1-forms, which are equivalent to the gravitational gauge potentials. The first and second Cartan Structure equations are as follows:
\begin{equation} \label{cartan_1}
T^a = D e^a = de^a+{\omega^a}_b \wedge e^b
\end{equation}
and
\begin{equation} \label{cartan_2}
R^{ab} = d \omega^{ab}+{\omega^a}_c \wedge \omega^{cb} \, ,
\end{equation}
where $T^a$ represents the torsion 2-forms, $R^{ab}$ denotes the curvature 2-forms, and $ d $ is the exterior derivative operator acting on the differential forms. The curvature 2-forms can also be expressed as
\begin{equation} \label{}
R^{ab}=\frac{1}{2} R^{ab}\,_{cd} e^c \wedge e^d \, ,
\end{equation}
where $ R^{ab}\,_{cd} $ denotes the components of the Riemann curvature tensor. Ricci 1-forms $P^a$ and the curvature scalar $R$ is obtained through the following contractions:
\begin{equation} \label{}
P^a = \iota_b R^{ba} = R^{ba}\,_{bc} \, e^c \equiv P^{a}\,_{c} \, e^c
\end{equation}
and
\begin{equation} \label{}
R = \iota_a P^a \, ,
\end{equation}
where $P^{a}\,_{c}$ denotes the components of the Ricci 1-forms.

\section{Action and variational procedure}

In the following section, we briefly outline the variational procedure leading to the field equations, emphasizing the role of quadratic curvature corrections within the $f(R)$ framework. We begin our analysis by revisiting the higher curvature gravity model, where gravitational action is a function of curvature scalar $R$ only. This model can be expressed as 
\begin{equation} \label{action}
I = \int_M\mathcal{L}[e, \omega]= \int_M f(R)\ast1 ,
\end{equation}
where $f(R)$ is the gravitational function given in terms of $R$. Under the assumption that the connections are torsion-free Levi-Civita connection, the Lagrangian density $n-$form of the theory can be written as 
\begin{equation} \label{torsion_free_Lag_density}
\int_M \bar{I}=\int_M \bar{\mathcal{L}}[e, \omega, \lambda] =\int_M \left( f (R) \ast1 + T^a \wedge \lambda_a \right) \, ,
\end{equation}
where $\{\lambda_a\}$ denotes the Lagrange multiplier $(n-2)$-forms. Using the abbreviation $e^{abc...}=e^a \wedge e^b \wedge e^c \wedge...$ for the wedge products and treating the connection as compatible and torsion-free unless otherwise specified, we calculate the arbitrary variations of the action density (\ref{torsion_free_Lag_density}) with respect to the co-frame 1-forms $e^a$, the connection 1-forms ${\omega^a}_b$ and the Lagrange multiplier $(n-2)$-forms $\lambda^a$ as
\begin{eqnarray} \label{}
\delta \bar{\mathcal{L}}[e, \omega, \lambda] &=& \delta e^c \wedge \left[ f(R) \ast e_c +f^\prime(R) \left(R^{ab} \wedge \ast e_{abc}-R \ast e_c\right) + D \lambda_c \right] \nonumber\\
&& + \delta \omega^{ab} \wedge \left[ D\left( f^\prime(R) \ast e_{ab}\right)+ \frac{1}{2} \left(e_b \wedge \lambda_a-e_a \wedge \lambda_b\right) \right] \nonumber\\
&& + \delta \lambda _a \wedge T^a + d \left[ f^\prime(R) (\delta \omega^{ab})\wedge \ast e_{ab} + (\delta e^a) \wedge \lambda_a \right] \, ,
\end{eqnarray}
where $\delta$ denotes variation in the corresponding field. For notational simplicity, we define $f^\prime(R) \equiv \frac{df(R)}{dR}$. By varying the action with respect to the dynamical fields:
\begin{equation} \label{}
\delta \bar{I} = 0 \, ,
\end{equation}
we obtain the following field equations:
\begin{equation} \label{ffe}
f(R)\ast e_c +f^\prime(R) \left(R^{ab} \wedge \ast e_{abc}-R \ast e_c\right)+ D \lambda_c =0
\end{equation}
and
\begin{equation} \label{sfe}
D \left( f^\prime(R) \ast e_{ab} \right) + \frac{1}{2} \left(e_b \wedge \lambda_a-e_a \wedge \lambda_b\right)=0 \, ,
\end{equation}
along with the automatically satisfied constraint relation
\begin{equation} \label{constraint}
T^a=d e^a+{\omega^a}_b \wedge e^b=0 \, .
\end{equation}
We rewrite the connection equation (\ref{sfe}) as
\begin{equation} \label{Lambda_def}
\Lambda_{ab} = 2 D\left( f^\prime(R) \ast e_{ab} \right) \, ,
\end{equation}
where
\begin{equation} \label{Lambda}
\Lambda_{ab}=e_a \wedge \lambda_b - e_b \wedge \lambda_a \, .
\end{equation}
By algebraically solving (\ref{Lambda}) for the Lagrange multiplier $(n-2)$-forms, we obtain
\begin{equation}
\lambda_b=2 \left( \iota^a D f^\prime(R)\right) \ast(e_{ab}) \, .
\end{equation}
This result indicates that the Lagrange multipliers become non-vanishing as a direct consequence of the $f(R)$ terms. In this context, these multipliers act as effective sources generated by higher-curvature corrections, ensuring the consistency of the torsion-free condition within the variational framework. Finally, the general form of the Einstein field equation becomes
\begin{equation} \label{efe_1}
f(R) \ast e_c +f^\prime(R) \left( R^{ab} \wedge \ast e_{abc} -R \ast e_c\right) + 2 D\left[  \left( \iota^a Df^\prime(R)\right) \ast e_{ac}  \right]=0 \, .
\end{equation}
The degeneracy of the field equations in this case reflects the reduced dynamical content of the theory under specific parameter choices, a feature that has also been observed in related higher-curvature gravity models. Additionally, we calculate the trace of the Einstein field equation for our model by multiplying (\ref{efe_1}) by $e^c$, yielding
\begin{equation}\label{trace}
\left(n f(R)-2Rf^\prime(R) \right) \ast 1-2(n-1)D \left( \iota^a D f^\prime(R)\right)\wedge \ast e_a=0 \, ,
\end{equation}
where $n$ represents the spacetime dimensions. Note that (\ref{efe_1}) and (\ref{trace}) are general expressions for $f(R)$ theories in the language of exterior algebra, and can be used to derive the field equations and trace for any $f(R)$ theory. Therefore, AdS solutions with constant curvature can be the solutions of $f(R)$ gravity models, both with and without torsion, because the trace equation admits solutions with constant scalar curvature.

In this work, we consider the following gravitational function:
\begin{equation}\label{gravitational_function}
f(R) = \frac{1}{2}  \left( \alpha R^2+R \right)+\Lambda \, ,
\end{equation}
where $\Lambda$ denotes the cosmological constant, and $\alpha$ represents the coupling constant. This functional form has been widely considered in literature as a simple extension of Einstein gravity including quadratic curvature corrections \cite{sotiriu}. The Einstein field equation and the trace of our model are given by
\begin{equation}\label{efe_2}
\left( \alpha R+\frac{1}{2}\right) R^{ab}\wedge \ast e_{abc} +\left( \Lambda-\frac{1}{2} \alpha R^2 \right)\ast e_c+2\alpha D \left(\iota_c \ast d R \right) = 0
\end{equation}
and
\begin{equation}
\left( n \left( \frac{1}{2}  \left( \alpha R^2+R \right)+\Lambda \right)-2R\left(\alpha R+ \frac{1}{2}\right)\right)\ast 1-2 \alpha (n-1) D \ast dR =0 \, .
\end{equation}
We observe that the theory admits topological Lifshitz-type solutions provided that the coupling constant $\alpha$, cosmological constant $\Lambda$ and curvature scalar $R$ satisfy the following relations: 
\begin{equation}\label{kosul}
\alpha=\frac{1}{8\Lambda} \, , \qquad R=-4\Lambda \, .
\end{equation}
These conditions are sufficient for the existence of the Lifshitz-type solutions considered herein. More general configurations with non-constant curvature may also be possible, although their explicit construction is beyond the scope of the present analysis. Finally, for the special value of the coupling constant given above, the gravitational function in action (\ref{action}) can be expressed as a perfect square:
\begin{equation}
\frac{1}{16 \Lambda}R^2+\frac{1}{2}R+\Lambda= \frac{ 1 }{ 16 \Lambda } (R+4\Lambda)^2 \, .
\end{equation}
Thus, the total gravitational action vanishes for $\alpha=1/8 \Lambda$ when the space-time curvature satisfies the condition $R=-4\Lambda$. We may then consider this solution or its Euclidean continuation as a type of gravitational instanton. This is related to the degeneracy of the field equation (\ref{efe_1}). If we disregard the case where $ f^\prime(R)\neq 0$, which leads to the Einstein equations with an effective cosmological constant and consequently excludes Lifshitz configurations, the only way for constant scalar curvature solutions to exist is if the scalar curvature value is a double root of the Lagrangian $f(R)$. Therefore, it is evident that the family of black holes derived in the following section will be solutions for any gravity theory described by the Lagrangian $f(R) = ( R + 4\Lambda)^2 h(R)$, where $h(R)$ is a function that remains regular at $R =- 4\Lambda$.

It is important to emphasize that all the solutions constructed in this work rely on a special parameter choice
\begin{equation} \label{con}
f^\prime (R)=1+2 \alpha R =0 \, ,
\end{equation}
which corresponds to the critical point of the $R^2$-corrected gravity theory. At this point, the effective gravitational coupling vanishes and the field equations become degenerate, allowing for a broad class of constant-curvature geometries that are not continuously connected to generic $f(R)$ dynamics. Such critical points are known to arise in higher-derivative gravity models and are associated with the disappearance of the standard gravity kinetic term in linearized theory \cite{deser}. For this reason, the spacetimes presented here should be viewed primarily as geometrical backgrounds admitted by the degenerate structure of the field equations at the critical point, rather than as generic dynamical solutions of $f(R)$ gravity. 

\section{Lifshitz-type solutions}

Similar Lifshitz-type solutions have been reported in the context of Einstein and higher-curvature gravities. Compared with these results, the solutions presented here incorporate quadratic curvature corrections and exhibit additional parameter dependence, which leads to a richer structure of possible geometries \cite{taylor}.
Lifshitz-type solutions in gravitational theories are significant because of their association with non-relativistic holography, functioning as gravitational duals (asymptotically Lifshitz spacetimes) for quantum field theories that demonstrate anisotropic scaling between space and time. This scaling is characterized by a dynamic critical exponent $z \neq 1$ meaning,
\begin{equation}
t \rightarrow \lambda^z t ,  \qquad \vec x \rightarrow \lambda \vec x.
\end{equation}
These solutions are generally categorized into two main families based on their time dependence: static and stationary solutions. Static solutions are time-independent and are often the first type of Lifshitz black holes found in various gravity models. A key characteristic is the existence of a global timelike Killing vector field that is perpendicular to the surfaces at constant time. This implies that the geometry remains unchanged when the time direction is reversed. On the other hand, stationary solutions possess a global timelike Killing vector field, which is not necessarily perpendicular to surfaces of constant time. For clarity, the analysis is organized into static and stationary subsections, while the thermodynamic properties and Lifshitz-type extremal solutions are presented separately in the other subsections.

\subsection{Lifshitz-type static solutions}

Einstein's equations can be solved in Lifshitz spacetimes, which exhibit specific scaling symmetry. These spacetimes are characterized by distinct scaling behaviors in space and time, leading to anisotropic scaling of the coordinates:
\begin{equation} \label{main_scaling}
t \rightarrow \lambda^z t , \qquad r \rightarrow \lambda^{-1} r , \qquad \vec x \rightarrow \lambda \vec x,
\end{equation}
where $z$ is the dynamic exponent, $\lambda$ is a constant and $\vec x$ is an $(n-2)$-dimensional vector. The metric that reflects the Lifshitz scaling behavior (\ref{main_scaling}) is given by
\begin{equation}
ds^2 =  -\left(\frac{r}{l}\right)^{2z} dt^{2}+\left(\frac{l}{r}\right)^2 dr^{2}+\left(\frac{r}{l}\right)^{2}d\bf{x}^2 \, ,
\end{equation}
where $l$ is the length scale $(l>0)$ and $d\bf{x}^2$$=\sum_{i=1}^{(n-2)}dx_i^2$. Here, we focus on solutions of the form $Li_{m}\times \Omega_{(n-m)}$, where $ Li_{m} $ represents $m$-dimensional Lifshitz-type submanifold and $ \Omega_{(n-m)} $ corresponds to an $ (n-m) $-dimensional subspace. By imposing scaling transformations:
\begin{equation} \label{321_1}
t \rightarrow \lambda^z t, \qquad r \rightarrow \lambda^{-1}r, \qquad \vec  y \rightarrow \lambda^\gamma \vec y, \qquad \vec x \rightarrow \lambda^N \vec x
\end{equation}
where $\vec y$ is an $(m-2)$-dimensional vector, $\vec x$ is an $(n-m)$-dimensional vector (if $ y_{m-2} \equiv \varphi $ is specified, then $ \vec y$ represents an $(m-3)$-dimensional vector for stationary spacetimes), and the Einstein field equation (\ref{efe_2}) yields the static vacuum solution (for $\kappa=0$) of the form $ Li_{m} \times \Omega_{(n-m)} $:
\begin{equation} \label{321_2}
ds^2 =  -\left(\frac{r}{l}\right)^{2z} dt^{2}+\left(\frac{l}{r}\right)^2 dr^{2}+\left(\frac{r}{l}\right)^{2\gamma} dy_{\alpha} dy^{\alpha}+\left(\frac{r}{l}\right)^{2N}dx_{i} dx^{i} \, ,
\end{equation}
under the conditions (\ref{kosul}) and
\begin{eqnarray} \label{321_4}
\Lambda &=& \frac{1}{4 l^2} \bigg( \left( N(n-m)+\gamma(m-2)\right)^2+2z\left(z+N(n-m)+\gamma(m-2)\right)\nonumber\\
&&+N^2(n-m)+\gamma^2(m-2) \bigg) \, .
\end{eqnarray}
Interestingly, as long as conditions (\ref{kosul}) and (\ref{321_4}) are satisfied, $R^2$-corrected gravity admits the following spacetime solution, which can also be identified as a black hole solution with spherical, hyperboloidal, or flat geometry:
\begin{equation} \label{321_5}
ds^2=-\left(\frac{r}{l}\right)^{2z} f^{2}(r) dt^{2}+\left(\frac{l}{r}\right)^2 \frac{dr^{2}}{f^{2}(r)}+\left(\frac{r}{l}\right)^{2 \gamma} dy_{\alpha} dy^{\alpha}+\left( \frac{r}{l}\right)^{2 N} \frac{dx_{i} dx^{i}}{\left(1+\kappa \frac{\rho^{2}}{4}\right)^{2}} \, ,
\end{equation}
where the metric function $f^2(r)$ is given by
\begin{equation} \label{321_6}
f^2(r)=1 +c_{2}\frac{l^{2N}}{r^{2N}} + c_{3}\frac{l^{p_{+}}}{r^{p_{+}}}+c_{4} \frac{l^{p_{-}}}{r^{p_{-}}} \, .
\end{equation}
Here,
\begin{eqnarray} \label{321_7}
p_{\pm}& = &\frac{1}{2}\Big(2 N (n-m)+3 z+2 \gamma (m-2)\nonumber\\
&& \pm \left.\sqrt{4 N (n-m)(z-N)+z^2+4\gamma(m-2)(z-\gamma)}\,\right)
\end{eqnarray}
and
\begin{equation} \label{321_8}
c_{2}=\frac{(n-m)(n-m-1)\kappa l^{2+2N}}{4 N^2-2 N d_1 + d_2}
\end{equation}
with
\begin{equation} \label{321_9}
d_{1}=2 N (n-m)+3 z+2 \gamma (m-2)
\end{equation}
and
\begin{eqnarray} \label{321_10}
d_{2}&=&N^2(n-m)(n-m+1)+2z^2+2z\gamma(m-2)+(m-2)(m-1)\gamma^2 \nonumber\\
&& +2N(n-m)\left( z+(m-2)\gamma\right)\, ,
\end{eqnarray}
where $ c_{3} $ and $ c_{4} $ are integration constants. In the solution, the parameter $z$ characterizes the anisotropic scaling symmetry of the Lifshitz spacetime, and $l$ represents the length scale inherent to the spacetime geometry. The curvature index $\kappa$ takes values of -1, 0, or +1, corresponding to the topology of the $(n-m)$-dimensional subspace (i.e. hyperbolic, planar (flat), and spherical geometry, respectively). Radius $\rho^{2}=x_{i}x^{i}$ defines the spatial dimensions of the $(n-m)$-dimensional subspace, illustrating how the spatial coordinates contribute to the metric structure. The index $\alpha$ ranges from 1 to $(m-2)$, indicating dimensions associated with specific properties of the solution, while the index $i$ ranges from $(m-3)$ to $(n-4)$, defining additional dimensions contributing to the overall geometry, provided $m<(n-1)$. For generic values of $z$ in $n$-dimensions, the $D$-dimensional solutions in references \cite{ayonbeato2} and \cite{sarioglu}  correspond to the specific case where $N=\gamma=1$. The function $f(R)$ must be real for the solution to describe a physically meaningful spacetime. This requires the following inequality to hold:
\begin{equation} \label{321_11}
4 N (n-m)(z-N)+z^2+4\gamma(m-2)(z-\gamma) \geq 0 \, .
\end{equation}
This inequality ensures the reality of the solution, which is crucial to its physical viability. The dynamic exponent $z$ must lie outside the interval $(z_{-}, z_{+})$, where
\begin{eqnarray} \label{321_12}
z_{\mp}&=&-2N(n-m)-2(m-2)\gamma \nonumber\\
&&\mp 2\sqrt{\left(N(n-m) + (m-2)\gamma \right)^2+N^2(n-m)+(m-2)\gamma^2}  \, .\nonumber\\
\end{eqnarray}
The boundaries $z_{-}$ and $z_{+}$ are expressed in terms of $N$, $n$, $m$, and $\gamma$, ensuring the consistency of the spacetime structure, particularly in relation to scaling behaviors relevant to physical applications. Solutions remain real for $z$ values outside the interval $(z_{-}, z_{+})$, meaning $z$ must be chosen appropriately (either greater than $z_{+} $ or less than $z_{-} $), to yield physically meaningful results. When selecting the branch $z > z_{+} > 0$ such that $p_+\geq p_- \geq z_+>0$, and choosing the integration constants $c_3$ and $c_4$ appropriately, the resulting solution can represent various geometries: {\it hyperbolic black hole} $(\kappa = -1)$: black hole with hyperbolic spatial geometry, {\it black brane} $(\kappa = 0)$: a high-dimensional black brane configuration; {\it spherical black hole} $(\kappa = +1)$: a black hole exhibiting spherical symmetry. In any case, the solution may exhibit a singularity at $r = r_+$, which corresponds to the largest root of $f^{2} ( r ) $ such that $ f^{2} (r_+ ) = 0 $. This point is identified as the outer horizon of the black hole, which is a critical feature of black hole thermodynamics. The horizon marks the boundary beyond which the gravitational pull becomes so intense that nothing, not even light, can escape, leading to considerations of the singularity at the core.

\noindent To ensure real-valued solutions, the parameters must be restricted such that the expression under the square root of (\ref{321_12}) remains non-negative. This imposed additional constraints on the allowed parameter space. Alternatively, inequality (\ref{321_11}) remains valid for generic values of scaling parameter $ z $ and parameter $ \gamma $ provided $ N $ is constrained within the interval $ N_{-} \leq N \leq N_{+} $. The boundaries of $N$ are defined as
\begin{equation}\label{321_13}
N_{\mp}=\frac{1}{2}\left(z \mp \sqrt{\frac{(n-m+1)z^2+4\gamma(m-2)(z-\gamma)}{(n-m)}} \,\right) \, .
\end{equation}
\noindent This ensures that the exponents $p_+$ and $p_-$ equal, thereby restoring the reality of the solution. Here, if we define
\begin{equation}
\gamma_{\mp}=\frac{1}{2} \left( z \mp \sqrt{\frac{(m-1)z^2-4Nz(n-m)-4N^2(n-m)}{(m-2)}} \right)
\end{equation}
$p_+$ again becomes equal to $p_-$ in the interval $\gamma_- \leq \gamma \leq \gamma_+$.

For the special case $ z= z_{ + } $, a logarithmic solution can be formulated as follows:
\begin{equation} \label{321_14}
f^2(r)= 1 + \bar{c}_{2}\frac{l^{2N}}{r^{2N}}+\frac{l^{p}}{r^{p}}\left(\bar{c}_{3}+\bar{c}_{4} \ln \left( \frac{r}{l}\right) \right) \, ,
\end{equation}
where
\begin{equation} \label{321_15}
p = \frac{1}{2}\Big(2 N (n-m)+3 z_{+} +2 \gamma (m-2) \Big) \, ,
\end{equation}
and
\begin{equation} \label{321_16}
\bar{c}_2\equiv c_2\mid_{z=z_+}\, .
\end{equation}
Interestingly, $ z = 0 $ is outside the defined intervals (i.e. $z\in (-\infty,z_{-}]\cup[z_{+},\infty)$). Nevertheless, solutions can still be derived for this special value of the scaling parameter $ z $ as follows:
\begin{equation} \label{321_17}
f^2(r)= 1 + \tilde{c}_2 \frac{l^{2N}}{r^{2N}} + \frac{l^{N(n-m)+\gamma(m-2)}}{r^{N(n-m)+\gamma(m-2)}} \left( \tilde{c}_3 \cos \left( \tilde{N} \ln \left( \frac{r}{l}\right)\right)+\tilde{c}_4 \sin \left( \tilde{N} \ln \left( \frac{r}{l}\right)\right) \right) \, ,
\end{equation}
where
\begin{equation} \label{321_18}
\tilde{N}=N^2(n-m)+\gamma^2(m-2)
\end{equation}
and
\begin{equation}\label{321_19}
\tilde{c}_2\equiv c_2\mid_{z=0} \, .
\end{equation}
\noindent This formulation demonstrates the flexibility of the solution space, even for values of $z$ that lie outside typical intervals, allowing for a broader exploration of spacetime geometries and their physical implications.

Additionally, it is worth noting that by expressing the spacetime solution (\ref{321_5}) in the form,
\begin{eqnarray} \label{321_34}
ds^2&=& \left( \frac{r}{l} \right)^{2 \theta} \left( - \left(\frac{r}{l}\right)^{2z^{\prime}} f^{2}(r) dt^{2}+\left(\frac{l}{r}\right)^2 \frac{dr^{2}}{\tilde{f}^{2}(r)}+\left(\frac{r}{l}\right)^{2} dy_{\alpha} dy^{\alpha} \right)\nonumber\\
&& +\left( \frac{r}{l}\right)^{2 N} \frac{dx_{i} dx^{i}}{\left(1+\kappa \frac{\rho^{2}}{4}\right)^{2}}
\end{eqnarray}
with
\begin{equation} \label{321_35}
\theta =  \gamma - 1 \, , \qquad  z^{\prime} =  z -  \gamma + 1 \, , \qquad  \tilde{f}^{2} (r) = f^{2} ( r )  \left( \frac{ r }{ l } \right)^{ 2 \gamma - 2 }  \, ,
\end{equation}
we obtain a hyperscaling violating $Li_{m} \times \Omega_{(n-m)} $ type static solution for the higher-dimensional $ R^{2} $-corrected gravity theory. 

\subsection{Lifshitz-type stationary solutions}

The Einstein field equation (\ref{efe_2}) can also accommodate stationary $Li_m \times \Omega_{(n-m)}$ type vacuum solutions (for $\kappa=0$), which take the form
\begin{eqnarray} \label{322_1}
ds^{2}&=&-\left(\frac{r}{l}\right)^{2z} dt^{2}+2\omega \left(\frac{r}{l}\right)^{z+\gamma} dt d\phi+\left(\frac{r}{l}\right)^{2\gamma} d\phi^2+\left(\frac{l^{2}}{r^{2}}\right) dr^{2} \nonumber\\
& & + \left(\frac{r}{l}\right)^{2\gamma} dy_{\alpha} dy^{\alpha}+\left(\frac{r}{l}\right)^{2N}dx_{i} dx^{i}
\end{eqnarray}
provided that the conditions (\ref{kosul}) and
\begin{eqnarray} \label{322_3}
\Lambda&=&\frac{2N^2(n-m)(n-m-1)-4N(z+\gamma(m-2))+2m\gamma^2(m-3)}{8l^2} \nonumber \\
&&+\frac{z^2(4+3\omega^2)+2z\gamma(2(m-2)+(2m-3)\omega^2)+\gamma^2(4+3\omega^2)}{8l^2(1+\omega^2)}\nonumber \\
\end{eqnarray}
\noindent are satisfied. Similar to the static case, the Einstein field equation (\ref{efe_2}) also allows for the following stationary $Li_m \times \Omega_{(n-m)}$ type spacetime solution (also referred to as the stationary black hole solution) if the relations (\ref{kosul}) and (\ref{322_3}) hold: 
\begin{eqnarray} \label{322_4}
ds^{2}&=&-\left(\frac{r}{l}\right)^{2z} f^{2}(r) dt^{2}+\left(\frac{r^2}{l^2}\right)\left(d\phi+\frac{\omega l^2}{r^2}dt\right)^2+\left(\frac{l^{2}}{r^{2}}\right) \frac{dr^{2}}{f^2(r)}\nonumber\\
& & + \left(\frac{r}{l}\right)^{2\gamma} dy_{\alpha} dy^{\alpha}+\left(\frac{r}{l}\right)^{2N}\frac{dx_{i} dx^{i}}{\left(1+\kappa \frac{\rho^{2}}{4}\right)^{2}} \, ,
\end{eqnarray}
where the metric function is given by:
\begin{equation} \label{322_5}
f^2(r)=1 +c_{2}\frac{l^{2N}}{r^{2N}} + c_{3}\frac{l^{p_{+}}}{r^{p_{+}}}+c_{4} \frac{l^{p_{-}}}{r^{p_{-}}} + c_{5} \frac{l^{2(z+1)}}{r^{2(z+1)}} \, .
\end{equation}
Here,
\begin{eqnarray}  \label{322_6}
p_{\pm}&=&\frac{1}{2}\Big(2+2N(n-m)+3z+2\gamma(m-3)\nonumber\\
&&\pm \sqrt{4(z-1)+4N(n-m)(z-N)+z^2+4\gamma (m-3)(z-\gamma)} \, \Big) \, , \nonumber\\
\end{eqnarray}
and
\begin{equation} \label{322_7}
c_{2}=\frac{\kappa (n-m) (n-m-1)\,l^{2+2N}}{4N^2-2Nd_{1}+d_{2}} \, ,
\end{equation}
\begin{equation} \label{322_8}
c_{5}=\frac{2 \omega^{2} l^{2+2z}}{4 (z+1)^2-2(z+1)d_1+d_2} \, ,
\end{equation}
together with
\begin{equation} \label{322_9}
d_{1}=2+2N(n-m)+3z+2\gamma(m-3) \, ,
\end{equation}
\begin{eqnarray} \label{322_10}
d_{2}&=& N^2(n-m)(n-m+1)+2(z^2+z+1)+2(m-3)(z+1)\gamma \nonumber\\
&&+(m-3)(m-2)\gamma^2+2N(n-m)\left(1+z+(m-3)\gamma \right) \, .
\end{eqnarray}
Here, $c_3$ and $c_4$ are integration constants. As previously mentioned, the curvature index $\kappa$ takes values of -1,0, or +1, corresponding to the hyperbolic, planar, and spherical geometries of the $(n-m)$-dimensional subspace. The stationary metrics presented in this section contain off-diagonal $dt d\phi$ terms, and are interpreted as rotating Lifshitz-type black hole configurations. It is worth noting that the rotation parameter $\omega$ cannot be eliminated by a global coordinate transformation owing to the periodic identification of the angular coordinate $\phi$, and therefore, represents a genuine physical parameter of the geometry. The existence of closed timelike curves is avoided, provided that the angular component $g_{\phi \phi}$ remains positive outside the horizon, which imposes mild constraints on the allowed range of parameters. Under these conditions, the Killing horizon is generated by the vector field $\xi=\partial_t + \Omega_H \partial_\phi$ where $\Omega_H$ denotes the angular velocity at the horizon as defined above. The geometries are thus regular in the domain outside the outer horizon and admit a consistent interpretation of rotating black hole spacetimes within the parameter ranges considered here. The geometries are thus regular in the domain outside the outer horizon and admit a consistent interpretation as rotating black hole spacetimes within the parameter ranges considered here, in the standard sense of Killing horizons \cite{wald}. The radius of this subspace is denoted by $\rho$, calculated as $\rho^2=x_{i}x^{i}$. For the stationary $Li_m \times \Omega_(n-m)$ type solutions, the spacetime index $\alpha$ ranges from 1 to $(m-3)$, whereas $i$ runs from $(m-2)$ to $(n-3)$. Parameter $z$ defines the anisotropic scaling symmetry of the Lifshitz spacetime, and $\omega$ denotes the rotation parameter. A physical solution requires that the metric function $f(r)$ be real, which can be ensured by the condition: 
\begin{equation} \label{322_11}
4(z-1)+4N(n-m)(z-N)+z^2+4\gamma (m-3)(z-\gamma) \geq 0 \, .
\end{equation}
From this, it is clear from (\ref{322_6}) that the dynamical exponent $z$ can take the values across the interval $z\in (-\infty,z_{-}]\cup[z_{+},\infty)$, where:
\begin{eqnarray}  \label{322_12}
z_{\mp}&=& -2-2N(n-m)-2\gamma(m-3) \nonumber \\
&& \mp 2\sqrt{1+N^2(n-m)+\gamma^2(m-3)+\left( 1+N(n-m)+\gamma(m-3)\right)^2} \, . \nonumber\\
\end{eqnarray}
The solution given by (\ref{322_4}) can represent a hyperbolic black hole for $\kappa=-1$, a spherical black hole for $\kappa=+1$, and a black brane for $\kappa=0$, assuming that the integration constants $c_1$ and $c_2$ are appropriately selected for $z>z_+>0$, ensuring that $p_+ \geq p_- \geq z_+ \geq 0$. In each case, $r_+$ corresponds to the (outer) horizon based on the assumption that the solution possesses a singularity at $r=r_+$, which indicates the largest root of $f^2(r)$ (i.e., $f^2(r_+)=0$). In addition, if we define
\begin{equation}\label{322_13}
N_\mp=\frac{1}{2}\left(z \mp \sqrt{\frac{\left(-4+(n-m+1)z^2-4\gamma^2(m-3)+4z(1+\gamma(m-3))\right)}{(n-m)}} \right)
\end{equation}
or
\begin{equation}
\gamma_{\mp}=\frac{1}{2} \left( z \mp \sqrt{\frac{((m-2)z+4)z+4Nz(n-m)+4N^2(n-m)-4}{(m-3)}}\right)
\end{equation}
$p_+$ again becomes equal to $p_-$.

When we set $ z= z_{ + } $, a logarithmic solution is obtained:
\begin{equation} \label{322_14}
\bar{f}^2(r)= 1 + \bar{c}_{2}\frac{l^{2N}}{r^{2N}}+\frac{l^{p}}{r^{p}}\left(\bar{c}_{3}+\bar{c}_{4} \ln \left( \frac{r}{l}\right) \right) + \bar{c}_{5} \frac{l^{2(z+1)}}{r^{2(z+1)}} \, ,
\end{equation}
where
\begin{equation} \label{322_15}
p = \frac{1}{2}\Big(2+2 N (n-m)+3 z_{+} +2 \gamma (m-3) \Big) \, ,
\end{equation}
and
\begin{equation} \label{322_16}
\bar{c}_2\equiv c_2\mid_{z=z_+}\, , \qquad \bar{c}_5\equiv c_5\mid_{z=z_+}\, .
\end{equation}
Moreover, it seems that no physical solution exists for $z=0$, as it falls outside the range $z\in (-\infty,z_{-}]\cup[z_{+},\infty)$. Nonetheless, if we analyze the field equation while accepting $z = 0$, we obtain a solution for this typical scaling parameter:

\begin{eqnarray} \label{322_17}
\bar{f}^2(r)&=& 1 + \tilde{c}_2 \frac{l^{2N}}{r^{2N}}+\tilde{c}_5 \frac{l^2}{r^2} \nonumber\\
&&+ \frac{l^{N(n-m)+\gamma(m-3)+1}}{r^{N(n-m)+\gamma(m-3)+1}} \left( \tilde{c}_3 \cos \left( \tilde{N} \ln \left( \frac{r}{l}\right)\right)+\tilde{c}_4 \sin \left( \tilde{N} \ln \left( \frac{r}{l}\right)\right) \right) \, , \nonumber\\
\end{eqnarray}
where
\begin{equation} \label{322_18}
\tilde{N}=N^2(n-m)+\gamma^2(m-3)+1
\end{equation}
and
\begin{equation} \label{322_19}
\tilde{c}_2\equiv c_2\mid_{z=0}\, , \qquad \tilde{c}_5\equiv c_5\mid_{z=0}\, .
\end{equation}
When the rotation parameter vanishes ($\omega=0$), stationary solutions are reduced to static solutions with an appropriate coordinate transformation. Moreover, by setting $N=1$ and $g_{0}=1/l$, the solutions were compatible with those presented in \cite{ayonbeato2} and \cite{sarioglu}. 

Furthermore, we note that by writing the spacetime solution (\ref{322_4}) as
\begin{eqnarray} \label{hv2}
ds^2&=& \left( \frac{r}{l} \right)^{2 \theta} \left( - \left(\frac{r}{l}\right)^{2z^{\prime}} f^{2}(r) dt^{2}+\left(\frac{r}{l}\right)^{2 \sigma}\left(d\phi+\frac{\omega l^2}{r^2} dt\right)^2+\left(\frac{l}{r}\right)^2 \frac{dr^{2}}{\tilde{f}^{2}(r)}\right.\nonumber\\
&& +\left.\left(\frac{r}{l}\right)^{2} dy_{\alpha} dy^{\alpha} \right)+\left( \frac{r}{l}\right)^{2 N} \frac{dx_{i} dx^{i}}{\left(1+\kappa \frac{\rho^{2}}{4}\right)^{2}}
\end{eqnarray}
with the following reparametrization
\begin{equation} 
\theta =  \gamma - 1 \, , \quad  z^{\prime} =  z -  \gamma + 1 \, , \quad  \sigma=2-\gamma \, , \quad \tilde{f}^{2} (r) = f^{2} ( r )  \left( \frac{ r }{ l } \right)^{ 2 \gamma - 2 }  \, .
\end{equation}
We obtained a hyperscaling violating $Li_{m} \times \Omega_{(n-m)} $ type stationary solution $R^2$-corrected gravity theory for generic dimensions. It is evident that when the rotation parameter vanishes ($\omega=0$), hyperscaling violating $Li_{m}\times\Omega_{(n-m)} $ type stationary solution (\ref{hv2}) reduces to hyperscaling violating $Li_{m}\times\Omega_{(n-m)} $ type static solution (\ref{321_34}) by taking $\sigma=1$ with an appropriate coordinate transformation. These solutions can be interpreted as conformally related to the original configurations. Therefore, they should be interpreted as belonging to the same class of geometries, rather than entirely independent solutions.

\subsection{Thermodynamic analysis}

At this stage, it is important to consider the thermodynamic characteristics associated with $Li_m \times \Omega_{(n-m)}$ type solutions. Even if a generic analytic expression for the horizon position $r_+$, which corresponds to the largest positive root of $f^{2}(r_{+})=0$, cannot be derived, thermodynamic analysis is still feasible. According to Wald's formula \cite{wald} for black hole entropy, we have

\begin{equation} \label{321_20}
S_{bh}=-2 \pi \int_{S} \frac{\delta \mathcal L}{\delta R_{abcd}} n_{ab} n_{cd} \, ,
\end{equation}
where $n_{ab}$ is the binormal to the horizon, defined as $n_{ab}=(\nabla_{a} \xi_{b}) / K$ and normalized so that $n_{ab}n^{ab}=-2$. The vanishing entropy in the $R^2$-corrected gravity action for these constant-curvature black hole solutions can be easily verified, given that satisfies $1+2\alpha R=0$. The vanishing of Wald entropy for the black hole solutions presented here follows directly from condition (\ref{con}), which is satisfied at the critical point of the theory. At this point, the usual definition of conserved charges and thermodynamic quantities in higher-curvature gravity theories becomes subtle because the effective gravitational coupling vanishes, and standard variational arguments underlying black hole thermodynamics no longer apply. In this sense, the coexistence of a finite Hawking temperature with vanishing entropy should not be interpreted as a conventional thermodynamic behavior but rather as a characteristic feature of black hole solutions arising at a degenerate point of the theory. Similar phenomena have been reported in other higher-derivative gravity models at critical points, where the standard notions of mass, entropy, and the first law require special care \cite{deser, sarioglu2}. 

\noindent The standard definition of surface gravity $K$ arises from the concept of Killing horizon, a hypersurface where the Killing vector $\xi$ of the metric becomes null. This surface gravity is expressed in terms of exterior forms as follows:

\begin{equation} \label{321_22}
K=\frac{1}{2}\left(\ast(d \xi \wedge \ast d \xi)\right)^{(1/2)} \, .
\end{equation}

\noindent For a static, spherically symmetric spacetime, the metric can be represented in coordinates as

\begin{equation} \label{321_23}
ds^{2}=-U(r)dt^{2}+\frac{dr^{2}}{V(r)}+R(r)d\Omega^{2} \, .
\end{equation}

\noindent Here, the (normalized) time-translational Killing vector for (\ref{321_23}) has components $(1,0,0,0,...)$. Surface gravity is determined by evaluating the limit of the metric properties near the horizon:

\begin{equation} \label{321_24}
K=\lim_{r\rightarrow r_{+}} \left(\frac{1}{2} \sqrt{\frac{V(r)}{U(r)}} U^{\prime}(r)\right) \, .
\end{equation}

\noindent The thermodynamic temperature of the black hole is directly related to this surface gravity and is given by:

\begin{equation} \label{321_25}
T=\frac{K}{2 \pi}.
\end{equation}

\noindent We can now compute the surface gravity and temperature for a static Lifshitz-type black hole by using the following equations:

\begin{equation} \label{321_26}
K=-\frac{\sqrt{n(n-1)}}{2}\lim_{r\rightarrow r_{+}}\left(\frac{r}{l}\right)^{z+1}\left(f^2(r)\right)^{\prime}
\end{equation}
\noindent and

\begin{equation} \label{321_27}
T=\frac{\sqrt{n(n-1)}}{4 \pi l} \left( 2 c_2 N \left(\frac{l}{{r_+}}\right)^{(2N-z)}+c_3 p_+ \left(\frac{l}{{r_+}}\right)^{(p_+-z)}+c_4 p_- \left(\frac{l}{{r_+}}\right)^{(p_--z)}\right) \, .
\end{equation}

\noindent In addition, the thermodynamic temperature of a stationary Lifshitz-type black hole  can be determined using the standard surface gravity concept $K$. The (normalized) time-translational Killing vector in the stationary case has components $(1,0,0,...,\Omega_H,0)$, where $\Omega_H$ represents the angular velocity, given by $\Omega_H=-\frac{g_{t \phi}}{g_{\phi \phi}}|_{r=r_+}=-\frac{\omega l^2}{{r_+}^2}$. From this, we derive the surface gravity and thermodynamic temperature of the Lifshitz-type stationary black hole as follows:

\begin{equation} \label{322_20}
K=-\frac{\sqrt{n(n-1)(n-2)}}{2}\lim_{r\rightarrow r_{+}}\left(\frac{r}{l}\right)^{z+1}\left(f^2(r)\right)^{\prime}
\end{equation}

\noindent and

\begin{eqnarray} \label{322_211}
T&=&\frac{\sqrt{n(n-1)(n-2)}}{4 \pi l} \bigg(2 c_2 N \left(\frac{l}{{r_+}}\right)^{(2N-z)}+c_3 p_+ \left(\frac{l}{{r_+}}\right)^{(p_+-z)}\nonumber\\
&&+c_4 p_- \left(\frac{l}{{r_+}}\right)^{(p_--z)}+2(z+1)c_5 \left(\frac{l}{{r_+}}\right)^{(z+2)} \bigg) \, .
\end{eqnarray}

\noindent The inclusion of the rotating parameter $\omega$ enriches the family of Lifshitz-type stationary black hole solutions and influences the temperatures calculated at the outer event horizon. Notably, the dimensions of the subspace also affects the temperature in both the static and stationary cases. Despite the vanishing entropy for these solutions, the temperatures remain non-zero because of the specific relation (\ref{kosul}) in $R^2$-corrected gravity theory. This property of Lifshitz black holes resemble that of BTZ black holes \cite{cai}. It is important to highlight that temperature equations (\ref{321_27}), and (\ref{322_211}) and the aforementioned solutions encompass a diverse range of solutions influenced by the curvature index $\kappa$, which determines whether the $(n-m)$-dimensional subspace is hyperbolic, flat, or spherical, corresponding to $\kappa=-1, 0, +1$. A comprehensive examination of the thermodynamics of Lifshitz-type black holes can be found in \cite{sarioglu2}.

The thermodynamic properties of the solutions can also be analyzed within the Wald Noether charge formalism \cite{wald, vollick}. In a diffeomorphism-invariant theory with Lagrangian $n$-form $\mathbf{L}$, the conserved current associated with vector field $\xi$ is given by
\begin{equation}
\mathbf{J}_\xi = \mathbf{\Theta}(\phi, \mathcal{L}_\xi \phi) - \iota_\xi \mathbf{L} \, ,
\end{equation}
which can be expressed on-shell as an exact form
\begin{equation}
\mathbf{J}_\xi = d \mathbf{Q}_\xi \, ,
\end{equation}
where $\mathbf{Q}_\xi$ is the Noether charge $(n-2)$-form. In the exterior algebra formalism, for $f(R)$ gravity presented here, the Noether charge takes the following form:
\begin{equation}
\mathbf{Q}_\xi = f'(R)\, \iota_\xi \omega^{ab} \, \ast e_{ab}+2 \iota_\xi \ast D f^\prime (R)\, .
\end{equation}
The conserved charges were obtained by integrating $\mathbf{Q}_\xi$ over a closed $(n-2)$-dimensional surface at spatial infinity. For the timelike Killing vector $\xi = \partial_t$, the energy is defined as
\begin{equation}
E = \int_{\partial\Sigma} \mathbf{Q}_{\partial_t},
\end{equation}
while for the rotational Killing vector $\xi = \partial_\phi$, the angular momentum is
\begin{equation}
J = \int_{\partial\Sigma} \mathbf{Q}_{\partial_\phi}.
\end{equation}
Because of the structure of $f(R)$ gravity, these charges acquire an overall factor of $f'(R)$, leading to rescaling relative to their Einstein gravity counterparts.

The first law of black hole thermodynamics follows from the variational identity
\begin{equation}
\delta \int_{\partial\Sigma} \mathbf{Q}_\xi = \delta E - \Omega_H \, \delta J,
\end{equation}
which, when evaluated at the horizon, yields
\begin{equation}\label{first_law}
\delta E = T \, \delta S + \Omega_H \, \delta J.
\end{equation}
Using the condition (3.18), one finds that
\begin{equation}
f'(R)=0,
\end{equation}
which implies that the Wald Noether charge vanishes identically,
\begin{equation}
\mathbf{Q}_\xi = 0.
\end{equation}
As a consequence, both the conserved energy and angular momentum vanish,
\begin{equation}
E = 0, \quad J = 0.
\end{equation}
Because Wald entropy also vanishes for the quadratic model (\ref{gravitational_function}) presented here, the first law of black hole thermodynamics (\ref{first_law}), is trivially satisfied. This behavior reflects the critical nature of the solutions, for which the effective gravitational coupling vanishes and all thermodynamic quantities degenerate simultaneously.

\subsection{Lifshitz-type extremal solutions}

We identified black hole solutions that demonstrate zero temperature under certain constant relations, referred to as extremal solutions. Initial examples of asymptotically Lifshitz black hole solutions with extremal horizons are presented in \cite{ayonbeato2}. Drawing inspiration from this work, we discovered a static Lifshitz-type extremal black hole solution for $\kappa=0$ based on the specific relationship between $c_3$ and $c_4$:
\begin{equation} \label{321_28}
c_3=-p_- \left(p_- - p_+\right)^{\frac{p_+ - p_-}{p_-}} \left( \frac{c_4}{p_+}\right)^{\frac{p_+}{p_-}} \, .
\end{equation}
This solution (i.e., (\ref{321_5})-(\ref{321_10})) has zero temperature and the extremal black hole is represented by:
\begin{eqnarray} \label{321_29}
ds^2& = & -\left(\frac{r}{l}\right)^{2z} \left(1-\frac{p_-}{p_- - p_+} \left( \frac{r_e}{r}\right)^{p_+}+\frac{p_+}{p_- - p_+} \left( \frac{r_e}{r}\right)^{p_-}\right) dt^{2}\nonumber\\
&& + \left(\frac{l}{r}\right)^2 \left(1-\frac{p_-}{p_- - p_+} \left( \frac{r_e}{r}\right)^{p_+}+\frac{p_+}{p_- - p_+} \left( \frac{r_e}{r}\right)^{p_-}\right)^{-1} dr^2  \nonumber\\
&&+ \left(\frac{r}{l}\right)^{2\gamma} dy_{\alpha} dy^{\alpha}+\left(\frac{r}{l}\right)^{2N}dx_{i} dx^{i} \, ,
\end{eqnarray}
where the extremal radius $r_e$ is
\begin{equation} \label{321_30}
r_e = l \left( \frac{p_- - p_+}{c_1 p_+}\, c_4 \right)^{1/2p_-} \, .
\end{equation}
Moreover, for $\kappa = \pm 1$, we discovered another extremal black hole solution, which was determined by the specific value of $z=z_+$. This solution arises from the following relation:
\begin{equation} \label{321_31}
c_3=-c_4- N \left( \frac{c_2}{p}\right)^{p/N} \left( N- p\right)^{(p-N)/N}
\end{equation}
and the corresponding spacetime geometry yields
\begin{eqnarray} \label{321_32}
ds^2& = & -\left(\frac{r}{l}\right)^{2z} \left(1+c_2 \left( \left( \frac{l}{r}\right)^{2N}-\frac{N}{p} \left( \frac{r_e}{r}\right)^{2p} \left( \frac{l}{r_e}\right)^{2N}\right)\right) dt^{2}\nonumber\\
&& + \left(\frac{l}{r}\right)^2 \left(1+c_2 \left( \left( \frac{l}{r}\right)^{2N}-\frac{N}{p} \left( \frac{r_e}{r}\right)^{2p} \left( \frac{l}{r_e}\right)^{2N}\right)\right)^{-1} dr^2  \nonumber\\
&&+ \left(\frac{r}{l}\right)^{2\gamma} dy_{\alpha} dy^{\alpha}+\left(\frac{r}{l}\right)^{2N}\frac{dx_{i} dx^{i}}{\left(1+\kappa \frac{\rho^{2}}{4}\right)^{2}} \, ,
\end{eqnarray}
where the extremal radius $r_e$ is defined as
\begin{equation} \label{321_33}
r_e=l\left( \frac{(N-p)}{p} \, c_2\right)^{1/2N} \, .
\end{equation}
Furthermore, we obtained a stationary Lifshitz-type extremal solution for $\kappa=0$, $p_-=(z+1)$ and under the following special relation:
\begin{equation} \label{322_21}
c_3=\frac{(z+1)}{p_+-(z+1)} \left( \frac{(z+1-p_+)(c_4+c_5)}{p_+}\right)^{p_+/(z+1)} \, .
\end{equation}
The corresponding spacetime geometry is:
\begin{eqnarray} \label{322_22}
ds^{2}&=&-\left(\frac{r}{l}\right)^{2z} \left( 1+(c_4+c_5) \left(\left( \frac{l}{r}\right)^{2(z+1)}-\frac{(z+1)}{p_+}\left(\frac{l}{r_e}\right)^{2(z+1)}\left( \frac{r_e}{r}\right)^{2p_+}\right)\right) dt^{2}\nonumber\\
&&+\left(\frac{r^2}{l^2}\right)\left(d\phi+\frac{\omega l^2}{r^2}dt\right)^2\nonumber\\
&&+\left(\frac{l^{2}}{r^{2}}\right) \left( 1+(c_4+c_5) \left(\left( \frac{l}{r}\right)^{2(z+1)}-\frac{(z+1)}{p_+}\left(\frac{l}{r_e}\right)^{2(z+1)}\left( \frac{r_e}{r}\right)^{2p_+}\right)\right)^{-1} dr^{2}\nonumber\\
&& + \left(\frac{r^{2}}{l^{2}}\right)^{2\gamma} dy_{\alpha} dy^{\alpha}+\left(\frac{r}{l}\right)^{2N}dx_{i} dx^{i} \, ,
\end{eqnarray}
where the extremal radius is:
\begin{equation} \label{322_23}
r_e=l\left( \frac{(z+1-p_+)(c_4+c_5)}{p_+}\right)^{1/(2z+2)} \, .
\end{equation}
Additionally, we found another black hole solution for $\kappa=\pm1$, for the specific value $z=z_+$, and under the special relation:
\begin{equation} \label{322_24}
c_3=-c_4-\frac{(z+1)}{(z+1-p)}\left( \frac{(z+1-p)(c_2+c_5)}{p}\right)^{p/(z+1)} \, .
\end{equation}
This extremal solution is given by:
\begin{eqnarray} \label{322_25}
ds^2&=&-\left(\frac{r}{l}\right)^{2z} \left( 1+(c_2+c_5)\left(\left( \frac{l}{r}\right)^{2(z+1)}-\frac{(z+1)}{p}\left(\frac{r_e}{r}\right)^{2p}\left( \frac{l}{r_e}\right)^{2(z+1)}\right)\right)dt^2 \nonumber\\
&&+\left(\frac{r^2}{l^2}\right)\left(d\phi+\frac{\omega l^2}{r^2}dt\right)^2\nonumber\\
&&+\left(\frac{l^{2}}{r^{2}}\right)\left( 1+(c_2+c_5)\left(\left( \frac{l}{r}\right)^{2(z+1)}-\frac{(z+1)}{p}\left(\frac{r_e}{r}\right)^{2p}\left( \frac{l}{r_e}\right)^{2(z+1)}\right)\right)dr^2 \nonumber\\
&& + \left(\frac{r^{2}}{l^{2}}\right)^{2\gamma} dy_{\alpha} dy^{\alpha}+\left(\frac{r}{l}\right)^{2N}\frac{dx_{i} dx^{i}}{\left(1+\kappa \frac{\rho^{2}}{4}\right)^{2}} \, ,
\end{eqnarray}
where the extremal radius is:
\begin{equation} \label{322_26}
r_e = l \left( \frac{(c_2+c_5)(z+1-p)}{p}\right)^{1/(2z+2)} \, .
\end{equation}

\section{Conclusion}

In this study, we explored $R^2$-corrected gravity theory through the Lagrangian density defined solely in terms of the curvature scalar $R$ and the cosmological constant. We formulated gravitational action and derived the associated field equations, which reveal the intricate relationships between the geometric properties of spacetime and the dynamics of gravity encapsulated in the $ f(R)$ formalism. We established that under specific conditions (i.e. $\alpha=\frac{1}{8\Lambda}$ and $R=-4\Lambda$), solutions of the Einstein field equations can yield a rich variety of spacetime geometries, including static and stationary Lifshitz-type black holes and branes with distinct topological structures of the type $Li_m \times \Omega_{(n-m)}$. The solutions obtained here are specific to the critical point of the theory and should be interpreted accordingly as geometries arising from the degenerate structure of the field equations. In this sense, the present results extend the existing Lifshitz black hole constructions by incorporating product-manifold geometries and hyperscaling violations within a single higher-curvature framework. Within this setting, the present analysis provides a unified higher-dimensional framework that simultaneously accommodates static, stationary and hyperscaling-violating Lifshitz-type geometries in $R^2$-corrected gravity.

Our analysis of new Lifshitz-type solutions emphasized the significance of the curvature index $ \kappa$ and the parameters $\gamma$ and $N$ (as is valid for the dynamical exponent $z$) in defining the geometry of the lower-dimensional subspaces, highlighting the intricate interplay between dimensionality and the thermodynamic properties of black holes. We have also provided new solutions, including extremal black hole configurations, and demonstrated that these spacetimes exhibit interesting thermodynamical behaviors where the entropy  vanishes for a non-zero horizon temperature. We revealed that the theory admits hyperscaling violating type static and stationary solutions for an arbitrary Lifshitz parameter $z$. These results suggest that the solutions correspond to a critical sector of the theory, characterized by vanishing conserved charges and entropy.

We note that as a future work, using a similar approach, one may investigate the product-manifold Lifshitz type black hole and black brane solutions in quadratic curvature gravity models. It would be interesting to extend the present construction beyond the critical point and to investigate whether similar classes of anisotropic solutions persist in non-degenerate higher-curvature gravity models.

\section{Acknowledgements}

I am deeply grateful to Prof. Dr. Hakan Cebeci for his invaluable guidance and critical feedback, which significantly enhanced the clarity and analytical depth of this study. I would also like to thank the anonymous reviewers for their valuable comments and suggestions, which helped to improve the quality of the paper. This work was supported by the Scientific Research Projects Commission of Eski\c{s}ehir Technical University under Grant No: 24GAP081.

\end{document}